\documentclass[aps,prx,reprint,superscriptaddress,groupedaddress]{revtex4-2}
\usepackage{amsmath}
\usepackage{amssymb}
\usepackage{amsthm}
\usepackage{graphicx}
\usepackage{physics}
\usepackage{bm}

\newtheorem{lemma}{Lemma}

\begin{document}

\title{Optimal control theory for measured quantum Schr\"odinger bridges
}
\author{Masayuki Ohzeki}
\email{mohzeki@tohoku.ac.jp}
\affiliation{Graduate School of Information Sciences, Tohoku University, Sendai 980-8579, Japan}
\affiliation{Department of Physics, Institute of Science Tokyo, Tokyo 152-8551, Japan}
\affiliation{Research and Education Institute for Semiconductors and Informatics, Kumamoto University, Kumamoto 860-8555, Japan}
\affiliation{Sigma-i Co., Ltd., Tokyo 108-0075, Japan}

\author{Andrew N. Jordan}
\affiliation{Institute for Quantum Studies, Chapman University, Orange, CA 92866, USA}
\affiliation{The Kennedy Chair in Physics, Chapman University, Orange, CA 92866, USA}
\affiliation{Department of Physics and Astronomy, University of Rochester, Rochester, NY, 14627, USA}

\date{\today}

\begin{abstract}
Schr\"odinger bridges and entropic optimal transport are usually formulated as stochastic interpolation problems between initial and final probability distributions.
In their computational form, the bridge potentials are obtained by Sinkhorn or iterative proportional fitting, and are often regarded as auxiliary scaling functions.
We show that for continuously monitored quantum systems, these potentials acquire a direct measurement-theoretic meaning.
The mathematical structure of conditional quantum trajectory theory induces a Fokker--Planck equation on quantum state space.
Conditioning this diffusion on a terminal distribution or a terminal measurement effect produces a Doob/Sinkhorn potential whose directional derivative along a unitary control vector field is the imaginary part of a generalized weak value.
The same construction connects the Schr\"odinger-bridge viewpoint to the optimal-path framework for continuously monitored trajectories: the backward bridge potential plays the role of an effect-like costate, and its weak-value directional derivative gives the local control signal.
By specifying the desired endpoint distribution, the induced drift produced by the Schr\"odinger bridge solution is the control solution that minimizes the quadratic cost feedback law, guiding the distribution to its desired endpoint.  We quantify the control score of the available control Hamiltonian as a logarithmic directional derivative of the bridge potential, or an imaginary weak value. Three explicit examples are given to illustrate the theory: an effect conditioned Doob bridge describing continuous collapse to a particular eigenstate, a full terminal-distribution Sinkhorn bridge mapping between quantum trajectory distributions, and two detectors measuring noncommuting observables, implementing the weak value/Sinkhorn-score drift control. This identifies weak measurement as a natural entropic regularization mechanism for quantum state transport and gives a route from Sinkhorn scaling to quantum feedback and Hamiltonian control synthesis for practical optimal control.
\end{abstract}

\maketitle

\section{Introduction}

Quantum control problems are often stated in terms of an initial preparation and a desired final condition \cite{guery2019shortcuts}.
For closed pure-state dynamics, this viewpoint leads to a two-boundary generator: the optimal coherent bridge between two rays is generated by the commutator of the corresponding projectors \cite{carlini2006time}.
In that setting, the imaginary weak value gives the local drift of the pre-selected state toward the post-selected boundary~\cite{Aharonov1988,DresselJordan2012,Dressel2014,Ohzeki2026TwoBoundary}.

When a quantum system is monitored continuously, the quantum state collapse rule is generalized to a continuous stochastic process that has been well developed theoretically \cite{JordanSiddiqi2024,Jacobs2014}. Continuous quantum measurement has now become routine to observe and investigate in experimental laboratories with the reconstructing of individual quantum trajectories~\cite{Murch2013,Weber2014}. Some examples of recent experimental quantum control in this area include reversing quantum jumps~\cite{Minev2019Nature}, and incorporating feedback for quantum information processing tasks, such as stabilizing Rabi oscillations~\cite{Rabi_vijay,patti2017linear}, or implementing continuous quantum error correction~\cite{QEC_livingston}.

A single measurement record produces a stochastic quantum trajectory, while the ensemble of records induces a diffusion on the space of quantum states.
This suggests that the appropriate extension of a coherent two-boundary bridge is not only a geodesic in Hilbert space, but a Schr\"odinger bridge on quantum state space.
The bridge is then conditioned either on a terminal probability distribution over states or on a terminal measurement event described by a positive effect.
This connects to trajectory-driving Hamiltonians and state-dependent Hamiltonian descriptions of collapse~\cite{HuJordan2023Driving,HuJordan2026Collapse}, and to stochastic path-integral and Pontryagin formulations of continuously monitored systems~\cite{ChantasriDresselJordan2013,weber2014mapping,chantasri2015stochastic,kokaew2026quantum,KarmakarJordan2026CDJPontryagin}.

Erwin Schr\"odinger posed the bridge problem to deal with unlikely events in classical statistical processes \cite{Schrodinger1931} - suppose the terminal distribution of a physical stochastic process is not the solution to the Fokker--Planck equation, but another probability distribution.  What then, is the interpolating bridging solution that has the correct marginals at the end points?
The classical Schr\"odinger bridge has a well-developed computational structure.
Given a reference Markov kernel, the optimal endpoint coupling has the multiplicatively scaled form $aKb$, and the scaling functions are obtained by Sinkhorn iteration \cite{sinkhorn1964relationship,georgiou2015positive} or iterative proportional fitting.
In continuous time, the same object appears as a Doob $h$-transform: the bridge drift is shifted by $D\nabla\log h$, where $D$ is the diffusion tensor.
The function $h$ is often introduced as a computational potential~\cite{Schrodinger1931,Leonard2014Survey,Sinkhorn1967,ChenGeorgiouPavon2016,PeyreCuturi2019}.
Quantum Schr\"odinger bridges have also been developed directly in terms of quantum Markovian dynamics and time-symmetric ensembles~\cite{MovillaSabbaghGeorgiou2025}.
Our formulation is complementary: we consider the diffusion induced on the manifold of conditioned quantum states by continuous measurement, and ask how the resulting Doob/Sinkhorn potentials are read out by physically available quantum controls.
Dressel and Jordan showed that the imaginary part of a generalized weak value is the logarithmic directional derivative of the post-selection probability along the unitary flow generated by the measured observable~\cite{DresselJordan2012}.
Here we use that response identity in a different setting: in a continuously monitored quantum system, the same derivative acts on the backward Doob/Sinkhorn potential of a state-space bridge.
When the backward potential is represented by an effect operator, its logarithmic derivative along a unitary control is twice the imaginary part of the corresponding generalized weak value.

This observation links four structures that are usually treated separately:
Schr\"odinger/Sinkhorn bridges \cite{Schrodinger1931}, Doob conditioning \cite{doob1984classical}, continuous quantum measurement \cite{JordanSiddiqi2024}, and weak values \cite{Aharonov1988,Dressel2014}.
It also connects naturally with recent trajectory-driving Hamiltonians \cite{carlini2006time,HuJordan2023Driving,HuJordan2026Collapse} and Pontryagin optimal-control formulations for monitored systems \cite{KarmakarJordan2026CDJPontryagin}.
The main message is that weak values do not merely decorate the bridge interpretation.
They are the observable directional derivatives of the Doob/Sinkhorn bridge potential along physically available control directions.

Several distinctions will be important below.
The Dressel--Jordan response formula concerns the logarithmic response of a post-selection probability; here the same derivative acts on the backward Doob/Sinkhorn potential generated by endpoint conditioning.
The CDJ-Pontryagin framework describes most-likely monitored trajectories and their costates \cite{chantasri2015stochastic,KarmakarJordan2026CDJPontryagin}, whereas the present bridge formulation uses an Eulerian backward potential on state space whose directional derivatives give local control scores.
Operator-valued quantum Schr\"odinger bridges construct bridges for quantum Markovian dynamics \cite{MovillaSabbaghGeorgiou2025}, while here the bridge is the ``classical'' diffusion induced on conditioned quantum states by continuous measurement within their Hilbert space.  We note that the effect operator formalism is also central to problems of quantum retrodiction as well as quantum smoothing, where both past and future information is used to make a refined estimate about the present state of the system \cite{guevara2015quantum,zhang2017prediction,garcia2017past}.  This is a continuous measurement version of the two-time reformulation of quantum mechanics \cite{aharonov1964time}.
Finally, trajectory-driving Hamiltonians prescribe a path and construct a Hamiltonian that follows it; in the present setting, the path ensemble itself is generated by the boundary condition through the Sinkhorn/Doob potential.

This article is organized as follows.  In Sec.~\ref{sec:boundary}, we introduce the past and future boundary conditions, and discuss the needed Hamiltonian to drive the system appropriately.  In Sec.~\ref{sec:bridge}, diffusive quantum trajectories are introduced, and the classical Schr\"odinger bridge problem is generalized to the continuous quantum measurement problem.  A physical interpretation of the bridge potential is given in terms of weak values and we discuss how the Sinkhorn drift can be directly applied to quantum control.  We also discuss how this formulation is closely connected to the stochastic action approach to quantum measurement and the most likely path between past and future boundary conditions.  In Sec.~\ref{sec:examples}, we give three physical examples of quantum Schr\"odinger bridges:  an effect-conditioned Doob bridge for continuous collapse to an eigenstate, a full terminal-distribution Sinkhorn bridge connecting distributions of quantum states, and a two-detector example of joint measurement of noncommuting observables implementing the weak-value/Sinkhorn-score drift as an explicit example of guiding a quantum system to its desired end distribution.
We give our conclusions and outlook for future work in Sec.~\ref{sec:outlook}.

\section{Past and future boundary conditions}
\label{sec:boundary}
In this section, we introduce the Hamiltonian flow and controls, diffusive process on quantum states of a measured system, the past and future boundary conditions, and the weak value response of an intermediate operation.  Let $\mathcal H$ be a finite-dimensional Hilbert space.
A quantum state is represented by a density matrix $\rho\ge0$ with $\Tr\rho=1$.
A pure state $\ket{\psi}$ corresponds to the rank-one projector $P_\psi=\ketbra{\psi}{\psi}$.
An effect is a positive operator $E$ satisfying $0\le E\le I$.
It represents a post-selection, a successful measurement event, or a terminal condition that is less specific than a final pure state.
The probability of this event in state $\rho$ is
\begin{equation}
    p_E(\rho)=\Tr(E\rho).
\end{equation}

A controllable Hamiltonian is written
\begin{equation}
    H(u)=H_0+\sum_{\mu=1}^m u_\mu A_\mu ,
\end{equation}
where $A_\mu=A_\mu^\dagger$ are available control generators and $u_\mu(t)$ are real control amplitudes.
The associated unitary perturbation generated by $A$ is
\begin{equation}
    \rho(g)=e^{-igA}\rho e^{igA}.
\end{equation}

We use $q=(q^1,\ldots,q^n)$ for coordinates on a state manifold.
For a qubit, for example, one may take $q=(x,y,z)$ and write
\begin{equation}
    \rho(q)=\frac{1}{2}\left(I+x\sigma_x+y\sigma_y+z\sigma_z\right).
\end{equation}
In these coordinates a continuously monitored state trajectory will be written as an It\^o stochastic differential equation
\begin{equation}
    dq^a=b^a(q,u)\,dt+\sum_\alpha B^a_\alpha(q)\,dW_t^\alpha ,
    \label{eq:sde}
\end{equation}
where $dW_t^\alpha$ are independent Wiener increments, $b^a$ is the drift, with diffusion tensor given by
\begin{equation}
    D^{ab}(q)=\sum_\alpha B^a_\alpha(q)B^b_\alpha(q).
\end{equation}
We will use two equivalent notations.
The expression $\rho(q)$ denotes the density matrix as a function on the state manifold, while $\rho_t$ or $\rho(t)$ denotes its value along a particular trajectory $q_t$:
\begin{equation}
    \rho(t)=\rho(q_t).
\end{equation}
Thus the Fokker--Planck and Sinkhorn equations are written in an Eulerian state-space representation, whereas the CDJ-Pontryagin equations are usually written along a trajectory.
Here CDJ-Pontryagin refers to the Chantasri--Dressel--Jordan stochastic path-integral framework \cite{chantasri2015stochastic} combined with Pontryagin optimal control for continuously monitored systems~\cite{lewalle2024optimal,KarmakarJordan2026CDJPontryagin}.
The two descriptions are related by evaluating the state-space fields at $q=q_t$.

For two pure boundary rays $\ket{\psi_i}$ and $\ket{\psi_f}$, define the boundary projectors by
\begin{equation}
    P_\nu=\ketbra{\psi_\nu}{\psi_\nu}.
\end{equation}
Here $\nu=i$ denotes the initial boundary and $\nu=f$ denotes the final boundary.
The coherent two-boundary bridge generator is proportional to
\begin{equation}
    i[P_f,P_i].
    \label{eq:closed-generator}
\end{equation}
This is Hermitian and therefore defines a physical Hamiltonian direction.
It is the generator that increases the overlap with the final boundary at the maximal local rate under the Hilbert--Schmidt quadratic cost \cite{carlini2006time}.
This closed-system generator is the starting point of the two-boundary optimal-control construction in Ref.~\cite{Ohzeki2026TwoBoundary}.
In the geodesic gauge $\braket{\psi_f}{\psi_i}=\cos\Theta>0$, with $\ket{\psi_f'}=(\ket{\psi_f}-\cos\Theta\,\ket{\psi_i})/\sin\Theta$, this commutator is proportional to the standard quantum-brachistochrone Hamiltonian $i\omega(\ketbra{\psi_f'}{\psi_i}-\ketbra{\psi_i}{\psi_f'})$; the remaining freedom is only the overall scale and the Hamiltonian gauge term proportional to the identity~\cite{carlini2006time}.

The same structure has an effect-boundary extension.

\begin{lemma}[Effect-boundary weak-value response]
Let $\rho$ be a density operator, let $E$ be an effect with $\Tr(E\rho)>0$, and let $A=A^\dagger$.
Define
\begin{equation}
    p_A(g)=\Tr\left(Ee^{-igA}\rho e^{igA}\right).
\end{equation}
Then
\begin{equation}
    \left.\frac{d}{dg}\log p_A(g)\right|_{g=0}
    =
    2\Im
    \frac{\Tr(EA\rho)}{\Tr(E\rho)} .
    \label{eq:weak-response}
\end{equation}
Equivalently,
\begin{equation}
    \left.\frac{dp_A(g)}{dg}\right|_{g=0}
    =
    \Tr\left(A\,i[E,\rho]\right).
    \label{eq:effect-generator}
\end{equation}
\end{lemma}

Equation~(\ref{eq:weak-response}) is the Dressel--Jordan logarithmic response formula~\cite{DresselJordan2012}, written for a mixed pre-selected state and an effect-valued post-selection.
It is the same response identity as Eq.~(41) of Ref.~\cite{DresselJordan2012}, expressed here in the notation needed for a terminal effect.
Equation~(\ref{eq:effect-generator}) shows that the local direction selected by the terminal effect is
\begin{equation}
    K_E(\rho)=i[E,\rho].
    \label{eq:effect-bridge-generator}
\end{equation}
If $\rho=P_i$ and $E=P_f$, this reduces to Eq.~(\ref{eq:closed-generator}).
Thus the closed coherent bridge is the rank-one special case of an effect-conditioned bridge.

\section{Continuous Measurement and the Fokker--Planck Bridge}
\label{sec:bridge}
We now lay out the mathematical structure of the diffusive process induced by continuous quantum measurement and give the quantum Schr\"odinger bridge construction.
Consider an efficient continuous measurement with measurement operators $L_\alpha$.
A standard conditional stochastic master equation can be written schematically as
\begin{equation}
    d\rho_t
    =
    \mathcal L_{u_t}(\rho_t)\,dt
    +
    \sum_\alpha \mathcal M_\alpha(\rho_t)\,dW_t^\alpha ,
    \label{eq:sme}
\end{equation}
where $\mathcal L_u$ includes Hamiltonian control, deterministic measurement back-action, and possible Lindblad terms, while $\mathcal M_\alpha$ gives the innovation part of the measurement back-action.
Here we focus on coherent control, but dissipation and measurement-based setting control is also possible \cite{kokaew2026quantum,lewalle2024optimal,KarmakarJordan2026CDJPontryagin} and has been implemented experimentally \cite{hacohen2018incoherent}.
In coordinates, this stochastic master equation becomes Eq.~(\ref{eq:sde}).
The probability density $p(q,t)$ of quantum trajectories then obeys the Fokker--Planck equation
\begin{equation}
    \partial_t p
    =
    -\partial_a\!\left[b^a(q,u)p\right]
    +
    \frac{1}{2}\partial_a\partial_b
    \!\left[D^{ab}(q)p\right].
    \label{eq:fp}
\end{equation}
This is the effective Fokker--Planck equation on quantum state space.
Its drift and diffusion are not externally imposed classical noise terms; they arise from measurement back-action, normalization of the conditioned state, and the chosen controls.
Equivalently, Eq.~(\ref{eq:fp}) may be symbolically written as 
\begin{equation}
    \partial_t p=\mathcal G_u^\ast p,
\end{equation}
where the forward generator acts on densities as
\begin{equation}
    \mathcal G_u^\ast p
    =
    -\partial_a\!\left[b^a(q,u)p\right]
    +
    \frac{1}{2}\partial_a\partial_b
    \!\left[D^{ab}(q)p\right].
    \label{eq:forward-generator}
\end{equation}

The adjoint operator $\mathcal G_u$ is the backward generator acting on test functions.
It is obtained by multiplying Eq.~(\ref{eq:forward-generator}) by a smooth test function $f(q)$, integrating over the state manifold, and moving the derivatives from $p$ to $f$ by integration by parts:
\begin{equation}
    \int f\,(\mathcal G_u^\ast p)\,dq
    =
    \int (\mathcal G_u f)\,p\,dq ,
\end{equation}
assuming boundary terms vanish or are absorbed into the boundary conditions.
This gives
\begin{equation}
    \mathcal G_u f
    =
    b^a(q,u)\partial_a f
    +
    \frac{1}{2}D^{ab}(q)\partial_a\partial_b f .
\end{equation}
For the uncontrolled reference process $u=0$, write $b_0=b(q,0)$ and $\mathcal G_0=\mathcal G$.

\subsection{Sinkhorn Potentials on Quantum State Space}
The classical Schr\"odinger bridge asks for the path measure $\pi^\ast$ closest in relative entropy to a reference diffusion while matching prescribed endpoint distributions $\mu_0$ and $\mu_T$.  We stress that because we consider a distribution of quantum trajectories in Hilbert space, there is no problem with the positivity of the marginals associated with pseudo-distributions of non-commuting phase space observables \cite{schleich2015quantum,arvidsson2024properties}.
In a discrete-time discretization with Markov kernel $K(q,q')$, the optimal endpoint coupling has the scaled form
\begin{equation}
    \pi^\ast(q,q')
    =
    a(q)K(q,q')b(q').
    \label{eq:sinkhorn-coupling}
\end{equation}
The functions $a$ and $b$ are obtained by Sinkhorn/IPFP updates.
Here IPFP denotes the iterative proportional fitting procedure (see e.g. Ref.~\cite{idel2016review} for a review), the alternating normalization of the two marginals:
\begin{equation}
    a^{(n+1)}(q)
    =
    \frac{\mu_0(q)}
    {\int K(q,q')b^{(n)}(q')\,dq'}.
\end{equation}
\begin{equation}
    b^{(n+1)}(q')
    =
    \frac{\mu_T(q')}
    {\int a^{(n+1)}(q)K(q,q')\,dq}.
\end{equation}
Here $\mu_0$ and $\mu_T$ are the prescribed endpoint probability densities, while $a^{(n)}$ and $b^{(n)}$ are the two scaling functions at the $n$th iteration.

In continuous time, the same structure is encoded in two positive bridge potentials,
\begin{equation}
    p^\ast(q,t)=\hat\varphi(q,t)\varphi(q,t),
\end{equation}
where $\hat\varphi$ is the forward bridge potential and $\varphi$ is the backward bridge potential.
The forward potential obeys
\begin{equation}
    \partial_t \hat\varphi=\mathcal G^\ast\hat\varphi.
\end{equation}
The backward potential obeys
\begin{equation}
    \partial_t \varphi=-\mathcal G\varphi.
\end{equation}
The initial boundary constraint is
\begin{equation}
    \hat\varphi(q,0)\varphi(q,0)=\mu_0(q).
\end{equation}
The final boundary constraint is
\begin{equation}
    \hat\varphi(q,T)\varphi(q,T)=\mu_T(q).
\end{equation}
This is a Schr\"odinger bridge for the diffusion induced on quantum state space.
It should be distinguished from operator-valued quantum Schr\"odinger bridges, where the forward and backward objects are matrices whose product represents a bridged density matrix~\cite{MovillaSabbaghGeorgiou2025}.
The corresponding Doob-transformed drift is
\begin{equation}
    b^\ast(q,t)
    =
    b_0(q)+D(q)\nabla\log\varphi(q,t).
    \label{eq:doob-drift}
\end{equation}
The bridge solution $p^\ast$ obeys the same Fokker-Planck equation (or corresponding stochastic Langevin equation) with this additional drift term.
The additional drift term is the mechanism by which the reference diffusion is steered to satisfy the endpoint constraints.
In the present setting, these constraints are the prescribed boundary distribution functions on quantum state space.
Here the vector notation means
\begin{equation}
    \left[D\nabla\log\varphi\right]^a
    =
    D^{ab}\partial_b\log\varphi .
\end{equation}

For a terminal effect rather than a full terminal distribution, the boundary condition is a terminal likelihood
\begin{equation}
    h_T(q)=\Tr\!\left[E_f\rho(q)\right].
    \label{eq:effect-terminal}
\end{equation}
Here $E_f$ is the terminal effect and $h_T(q)$ is the probability of that event when the system is in the state $\rho(q)$.
The backward function $h(q,t)$ obeys
\begin{equation}
    \partial_t h+\mathcal G h=0,
    \label{eq:h-backward}
\end{equation}
with terminal condition $h(q,T)=h_T(q)$.
The conditioned process is the Doob transform with
\begin{equation}
    b^E(q,t)=b_0(q)+D(q)\nabla\log h(q,t).
    \label{eq:effect-doob}
\end{equation}
Thus, a full endpoint-distribution problem gives a genuine Sinkhorn bridge, while a terminal effect gives an effect-conditioned Doob bridge.
Both use the same backward potential structure.

\subsection{Physical Interpretation of the Bridge Potential}
A natural question that arises is if there is any physical meaning of the potentials $\varphi(q, t), {\hat \varphi}(q,t)$, or if they only play an abstract and auxiliary role in the theory.
To answer this question, let $f_A(q)$ be the vector field on state space generated by the Hamiltonian $A$:
\begin{equation}
    \left.\frac{d}{dg}\rho(q_g)\right|_{g=0}
    =
    -i[A,\rho(q)] .
\end{equation}
For any positive bridge potential $\varphi(q,t)$, define the control score
\begin{equation}
    S_A(q,t)
    =
    f_A(q)\cdot\nabla\log\varphi(q,t).
    \label{eq:general-score}
\end{equation}
This is the directional derivative of the Sinkhorn/Doob potential along the unitary control vector field.
Here $f_A$ is the vector field along the Hamiltonian direction $A$: if $\Phi_g^A(q)$ denotes the state-space flow induced by $e^{-igA}\rho(q)e^{igA}$, then
\begin{equation}
    f_A^a(q)=\left.\frac{d}{dg}\Phi_g^{A,a}(q)\right|_{g=0}.
\end{equation}

When the potential is represented by a backward effect, we then have
\begin{equation}
    \varphi(q,t)=\Tr[E(t)\rho(q)],
    \label{eq:affine-potential}
\end{equation}
where $E(t)$ is the backward effect at time $t$.
Differentiating this likelihood along the unitary flow generated by $A$ gives
\begin{equation}
    S_A(q,t)
    =
    2\Im
    \frac{\Tr[E(t)A\rho(q)]}{\Tr[E(t)\rho(q)]}.
    \label{eq:sinkhorn-weak}
\end{equation}
The generalized weak value associated with the pair $(E(t),\rho(q))$ is
\begin{equation}
    (A)_w^{E,\rho}
    =
    \frac{\Tr[E(t)A\rho(q)]}{\Tr[E(t)\rho(q)]}.
\end{equation}
Therefore $S_A(q,t)=2\Im(A)_w^{E,\rho}$.

This is the proposed physical meaning of the bridge potential in quantum state space.
The potential is not merely a numerical scaling function.
It is the backward likelihood of the terminal event, and its response to a physical Hamiltonian perturbation is the imaginary weak value.

\subsection{From Sinkhorn Drift to Quantum Control}
The solution of the quantum Schr\"odinger
bridge problem gives the drift needed to account for the final (observed) boundary condition, that may not be the solution to the original diffusion equation, resulting from the possibility of rare events.   However, this solution can be turned around and reapplied:  By exercising the control variables at hand, the system may be actively guided to the desired final boundary condition as a {\it quantum control problem}.  
In the ideal case, the Doob drift correction to the Fokker--Planck equation can be reproduced by the available control terms to make the best-case control.

However, the ideal bridge drift in Eq.~(\ref{eq:doob-drift}) or Eq.~(\ref{eq:effect-doob}) need not coincide with an available quantum control.
The available Hamiltonians generate vector fields $f_\mu(q)$, and measurement settings may generate additional controllable drift and diffusion directions.
The control synthesis problem is therefore a projection:
\begin{equation}
    D(q)\nabla\log\varphi(q,t)
    \longrightarrow
    \mathrm{span}\{f_\mu(q)\}.
\end{equation}
This equation is not yet a control protocol; it identifies the ideal entropic-bridge drift that the physical controls should approximate.
For a quadratic control penalty, a local feedback law can be written as
\begin{equation}
    u^\ast(q,t)
    =
    \arg\min_u
    \left\|
    D\nabla\log\varphi-\sum_\mu u_\mu f_\mu
    \right\|_M^2
    +
    \lambda \sum_\mu u_\mu^2 ,
    \label{eq:projection}
\end{equation}
where $M$ is a chosen metric or pseudometric on the tangent space of the state manifold.
Equivalently, the coefficients $u_\mu$ are determined by the normal equations
\begin{equation}
C_\mu =   \sum_\nu
    \left(
    \langle f_\mu,f_\nu\rangle_M+\lambda\delta_{\mu\nu}
    \right)u^\ast_\nu
    =
    \left\langle f_\mu,D\nabla\log\varphi\right\rangle_M .
    \label{eq:projection-normal}
\end{equation}
The right-hand side is the bridge demand seen along the available control direction $f_\mu$.
This is the first meaning of a control score $C_\mu$: it ranks how strongly each allowed control contributes to reproducing the Sinkhorn drift.  This quantity should be distinguished from the score definition for $S_\mu$ given in Eq.~(\ref{eq:general-score}), the quantity directly connected with the imaginary weak value. This is the ``weak-value score'' in the sense of sensitivity of the boundary likelihood along the control vector field, whereas the control score $C_\mu$ in (\ref{eq:projection-normal}) also accounts for the diffusion tensor, and is directly applicable to the quantum control problem when implementing the bridge drift.

This terminology is also related to the score in estimation theory.
If a small control displacement $g$ changes the backward potential as
\begin{equation}
    \varphi_g(q,t)=\varphi(\Phi_g^{A_\mu}(q),t),
\end{equation}
and if one forms the normalized tilted ensemble
\begin{equation}
    P_g(q,t)=
    \frac{p(q,t)\varphi_g(q,t)}
    {\int p(q,t)\varphi_g(q,t)\,dq},
\end{equation}
then
\begin{equation}
    \left.\partial_g\log P_g(q,t)\right|_{g=0}
    =
    S_\mu(q,t)-\mathbb E_{P_0}[S_\mu].
\end{equation}
The Fisher information for this local control displacement is therefore
\begin{equation}
    I_\mu(t)=
    \mathbb E_{P_0}
    \left[
    \left(S_\mu-\mathbb E_{P_0}[S_\mu]\right)^2
    \right].
\end{equation}
Thus the expected square of the bridge score controls the sensitivity of the conditioned state ensemble to the control direction, up to the subtraction of its mean response.

There is a second, more intrinsic score when the immediate objective is to increase the terminal likelihood rather than to reproduce the whole drift vector.
For a Hamiltonian direction $A_\mu$, define the score as the logarithmic directional derivative of the bridge potential:
\begin{equation}
    S_\mu(q,t)
    =
    f_\mu(q)\cdot\nabla\log\varphi(q,t).
    \label{eq:directional-score}
\end{equation}
When the potential is represented by a backward effect, this score becomes
\begin{equation}
    S_\mu(q,t)
    =
    2\Im
    \frac{\Tr[E(t)A_\mu\rho(q)]}{\Tr[E(t)\rho(q)]}.
    \label{eq:switching}
\end{equation}
This is the weak-value form of the bridge control rule.

The situation is directly analogous to the synthesis of closed-system algorithms.
There, the ideal generator is $i[P_f,P_i]$, and the circuit problem is to synthesize or project this generator into the available gate algebra.
Here, the ideal stochastic drift is $D\nabla\log\varphi$, and the feedback-control problem is to synthesize or project it into the available Hamiltonian and measurement-control vector fields.

\subsection{Connection to CDJ-Pontryagin Optimal Control}

The Chantasri--Dressel--Jordan path-integral formulation assigns a stochastic action to measurement records and quantum trajectories \cite{ChantasriDresselJordan2013,chantasri2015stochastic,JordanSiddiqi2024}.
Recent Pontryagin formulations introduce a costate operator $\sigma(t)$ and a Pontryagin Hamiltonian of the form
\begin{equation}
    \mathcal K(\sigma,\rho,r,u)
    =
    \Tr[\sigma F(\rho,r,u)]
    +
    G(\rho,r),
    \label{eq:pontryagin-hamiltonian}
\end{equation}
where $F$ is the conditional state update equation and $G$ is the log-likelihood rate of the measurement record \cite{KarmakarJordan2026CDJPontryagin}.
Here $\rho$ denotes the trajectory value $\rho(t)=\rho(q_t)$, not a different object from the state-space function $\rho(q)$ used above.
In the same way, an effect-like costate $\sigma(t)$ may be viewed as the backward boundary field evaluated along the same trajectory.
The stationarity conditions for the stochastic action 
\begin{equation}
    {\cal S} = \int dt ( -{\rm Tr}[\sigma {\dot \rho}] +  \mathcal K),
\end{equation}
first give the forward state equation
\begin{equation}
    \dot\rho=F(\rho,r,u).
    \label{eq:cdj-forward-state}
\end{equation}
They also give the backward costate equation
\begin{equation}
    \dot\sigma
    =
    -\left[\partial_\rho F(\rho,r,u)\right]^\dagger[\sigma]
    -\partial_\rho G(\rho,r).
    \label{eq:cdj-backward-costate}
\end{equation}
In the most-likely-path conception of Ref.~\cite{KarmakarJordan2026CDJPontryagin}, the most-likely readout $\bar r$ is inserted into the action as part of the extremization and the normalization $\Tr(\sigma\rho)=1$ is chosen.
With this choice, corresponding to Eq.~(14) of Ref.~\cite{KarmakarJordan2026CDJPontryagin}, the explicit likelihood-gradient term in Eq.~(\ref{eq:cdj-backward-costate}) drops out and the costate equation becomes
\begin{equation}
    \dot\sigma=-F^\dagger(\sigma,{\bar r},u),
\end{equation}
which is the time-reversed evolution expected for an effect-like variable.
For a Hamiltonian control contribution
\begin{equation}
    F_u(\rho)=-i[H(u),\rho],
\end{equation}
the {\it switching function} measures the first-order change of the Pontryagin Hamiltonian with respect to the control.
In bang-bang control, its sign determines which extreme value of the control is selected \cite{boscain2021introduction}.
For the Hamiltonian contribution, the first trace form is
\begin{equation}
    \frac{\partial\mathcal K}{\partial u_\mu}
    =
    -i\Tr\!\left(\sigma[A_\mu,\rho]\right).
\end{equation}
Using cyclicity of the trace and Hermiticity gives
\begin{equation}
    \frac{\partial\mathcal K}{\partial u_\mu}
    =
    2\Im\Tr(\sigma A_\mu\rho).
    \label{eq:cdj-switch}
\end{equation}
If the costate is normalized as a backward effect, this becomes
\begin{equation}
    \frac{1}{\Tr(\sigma\rho)}
    \frac{\partial\mathcal K}{\partial u_\mu}
    =
    2\Im
    \frac{\Tr[\sigma(t)A_\mu\rho(t)]}{\Tr[\sigma(t)\rho(t)]}.
\end{equation}
Thus the CDJ-Pontryagin switching function is the same object as the unitary directional derivative of the Sinkhorn/Doob potential whenever $\sigma(t)$ represents the backward boundary likelihood.

This also clarifies the limitations of the effect language.
For terminal post-selection probabilities, $\sigma(t)$ is naturally an effect-like backward operator.
For more general endpoint constraints or arbitrary cost functionals, the costate is a Lagrange multiplier and need not be positive.
In those cases, Eq.~(\ref{eq:cdj-switch}) still provides the Pontryagin control score, while Eq.~(\ref{eq:sinkhorn-weak}) is recovered when the costate is positive and admits an effect interpretation.

\section{Physical examples of quantum Schr\"odinger bridges}
\label{sec:examples}
Three examples described in this section illustrate the above constructions: an effect-conditioned qubit Doob bridge, a full terminal-distribution Sinkhorn bridge, and a two-detector qubit bridge implementing the weak-value/Sinkhorn-score drift.  In the last example, a numerical illustration will be given of the quantum control aspect of the solution.

\subsection{Continuously Measured Qubit}
\begin{figure*}
    \centering
    \includegraphics[width=\linewidth]{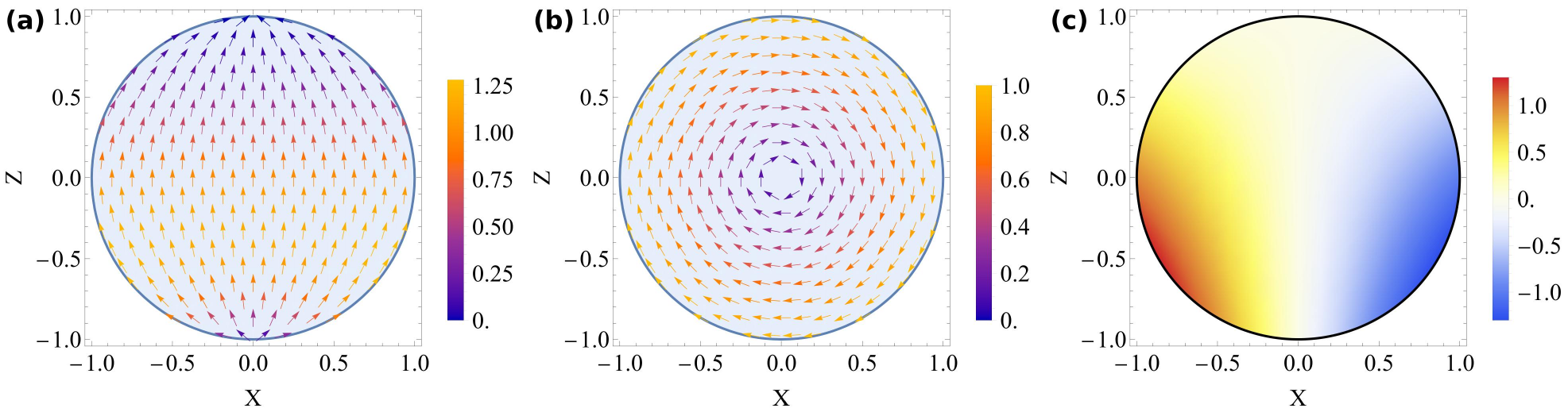}
    \caption{Illustration of the effect-conditioned Doob bridge on the XZ slice of the Bloch ball:  Panel (a) The Doob corrected drift $D\nabla\log h$  is plotted as a vector field.  The physical drift shows an intuitive mapping of the state space to the terminal collapse state $|0\rangle \langle 0|$. 
    Panel (b) is the control vector field $f_y(q)$.  Panel (c) is the diffusion compensated score for $y$-control, $C_y = \langle f_y(q) , D\nabla\log h \rangle_M = x(z-1)$.}
    \label{fig:Doob-bridge}
\end{figure*}
For a qubit continuously measured in the $\sigma_z$ basis with strength $\kappa$, we write a parameterized density matrix
\begin{equation}
    \rho(q)=\frac{1}{2}(I+x\sigma_x+y\sigma_y+z\sigma_z).
\end{equation}
With a $y$-control Hamiltonian $H(u)=u\sigma_y/2$, the efficient-measurement stochastic master equation is
\begin{equation}
    d\rho
    =
    -i[H(u),\rho]\,dt
    +
    \kappa\mathcal D[\sigma_z]\rho\,dt
    +
    \sqrt{\kappa}\mathcal H[\sigma_z]\rho\,dW_t,
    \label{eq:qubit-sme}
\end{equation}
where
\begin{equation}
    \mathcal D[L]\rho=L\rho L^\dagger-\frac{1}{2}\{L^\dagger L,\rho\}.
\end{equation}
The innovation superoperator in the same equation is
\begin{equation}
    \mathcal H[L]\rho=L\rho+\rho L^\dagger-\Tr[(L+L^\dagger)\rho]\rho .
\end{equation}
Here we are measuring in the $z$-basis, so the Lindblad operator is $L = \sigma_z$ for our example.  This convention follows the standard efficient-measurement stochastic master equation; we note that the characteristic measurement time is $\tau_M = 1/(4 \kappa)$ in this notation \cite{JordanSiddiqi2024}.
In Bloch coordinates, Eq.~(\ref{eq:qubit-sme}) gives three It\^o stochastic differential equations.
The $x$ component is
\begin{equation}
    dx=(uz-2\kappa x)\,dt-2\sqrt{\kappa}\,xz\,dW_t.
    \label{eq:qubit-bloch-sde-x}
\end{equation}
The $y$ component is
\begin{equation}
    dy=-2\kappa y\,dt-2\sqrt{\kappa}\,yz\,dW_t.
    \label{eq:qubit-bloch-sde-y}
\end{equation}
The $z$ component is
\begin{equation}
    dz=-ux\,dt+2\sqrt{\kappa}(1-z^2)\,dW_t.
    \label{eq:qubit-bloch-sde-z}
\end{equation}
Thus the control vector field generated by $A=\sigma_y/2$ is given by a column vector,
\begin{equation}
    f_y(q)=
    \begin{pmatrix}
        z\\
        0\\
        -x
    \end{pmatrix},
    \label{eq:fy}
\end{equation}
up to the sign convention in $H=uA$.
The measurement-induced diffusion is anisotropic and state dependent.
The stochastic term, which is independent of the Hamiltonian control, is given by
\begin{equation}
    B(q)
    =
    2\sqrt{\kappa}
    \begin{pmatrix}
        -xz\\
        -yz\\
        1-z^2
    \end{pmatrix},
\end{equation}
so the diffusion tensor $D=BB^\top$ is rank one.
Taking the reference drift with no Hamiltonian control, the backward generator is therefore
\begin{equation}
    \mathcal G f
    =
    -2\kappa x\,\partial_x f
    -2\kappa y\,\partial_y f
    +\frac{1}{2}B^a(q)B^b(q)\partial_a\partial_b f .
    \label{eq:qubit-backward-generator}
\end{equation}

Now choose the terminal effect
\begin{equation}
    E_f=\ketbra{0}{0}=\frac{1}{2}(I+\sigma_z),
\end{equation}
representing the case when quantum state collapse is onto state $|0\rangle$.
The terminal likelihood is
\begin{equation}
    h_T(q)=\Tr[E_f\rho(q)]=\frac{1+z}{2},
\end{equation}
showing that if the initial state is $|1\rangle$, corresponding to $z=-1$, then there is no possibility to end in the orthogonal state, while if the initial state is $|0\rangle$, corresponding to $z=1$, then there is unit probability to end in the selected final state.
Since $\mathcal G h_T=0$, the backward equation
$\partial_t h+\mathcal G h=0$ gives the time-independent solution
\begin{equation}
    h(q,t)=\frac{1+z}{2}.
    \label{eq:qubit-h}
\end{equation}
The logarithmic directional derivative of this Doob potential along the $y$-control is
\begin{equation}
    S_y(q,t)
    =
    -\frac{x}{1+z}.
    \label{eq:qubit-score}
\end{equation}

Here $S_y(q,t)$ is the quantity $f_y(q)\cdot\nabla\log h(q,t)$ evaluated for Eq.~(\ref{eq:qubit-h}).
The same number is obtained directly from the generalized weak value:
\begin{equation}
    2\Im
    \frac{\Tr[E_f(\sigma_y/2)\rho(q)]}{\Tr[E_f\rho(q)]}
    =
    -\frac{x}{1+z}.
\end{equation}
Note that both the generalized weak value and gradient of the Doob potential diverges when $z\rightarrow -1$, corresponding to state $|1\rangle$.  This reflects the vanishing overlap between the initial and final conditions, which is the characteristic weak value divergence \cite{Dressel2014}.   This example is the effect-conditioned Doob case rather than a full endpoint-distribution Sinkhorn bridge.
It shows explicitly that the backward likelihood $h$ is the physical bridge potential and that its Hamiltonian directional derivative is the imaginary weak value.

The Hamiltonian score in Eq.~(\ref{eq:qubit-score}) tells how an available unitary control locally increases the same terminal likelihood.  
The full Doob drift $D\nabla\log h$ is generated by conditioning the measurement diffusion, resulting in the drift vector
\begin{equation}
 b^\ast(x,y,z) = b_0 +  4 \kappa   \begin{pmatrix} -x z (1-z) \\
 - y z (1-z) \\
 (1-z)^2 (1+z)
 \end{pmatrix}.
\end{equation}
The Doob drift is plotted in Fig.~\ref{fig:Doob-bridge}(a), showing a smooth vector field taking every point in the $x-z$ slice of the Bloch sphere to the collapse state $|0\rangle\langle 0|$.  The diffusion tensor regularizes the divergence in the gradient of $h$.
Fig.~\ref{fig:Doob-bridge}(b) plots the available control vector field, and panel (c) plots the diffusion-compensated score, $C_y = \langle f_y(q) , D\nabla\log h \rangle_M = x(z-1)$, relevant for implementing the control strategy.  As mentioned in the definition of the control score $C_\mu$, given in Eq.~(\ref{eq:projection}), the inner product is relative to a metric $M$. Here, we have taken the flat metric on the Bloch ball. Another natural choice is the Bures metric \cite{bures1969extension} for the Bloch ball \cite{hubner1992explicit}. However, this metric introduces an additional geometric choice and has a radial singularity at the pure-state boundary. In the present Hamiltonian-control example, the control vector field is tangent to the radial direction of the Bloch ball, so the singular radial term in the Bures metric does not contribute to this particular overlap. Thus, the Bures inner product gives essentially the same control score up to an overall factor.

\subsection{Analytic Sinkhorn Bridge with a Full Terminal Distribution}

A minimal analytic example treats a terminal condition given by a full distribution $\mu_T(q)$, not a single effect.
To make the control direction explicit, take $q$ to be an unwrapped meridian angle of a qubit.
The corresponding state is
\begin{equation}
    \rho(q)=
    \frac{1}{2}
    \left(I+\sin q\,\sigma_x+\cos q\,\sigma_z\right).
    \label{eq:meridian-qubit-chart}
\end{equation}
Here $q$ is generated by rotation about the $y$ axis from $\ketbra{0}{0}$.
A chart means a local coordinate patch on the state manifold in which nearby density matrices are parametrized by ordinary coordinates.
The bridge is formulated in this local coordinate rather than globally on the Bloch sphere.
The Hamiltonian generator $A=\sigma_y/2$ translates this coordinate:
\begin{equation}
    e^{-igA}\rho(q)e^{igA}=\rho(q+g).
\end{equation}
The associated vector field on the coordinate $q$ is
\begin{equation}
    f_A(q)=1.
    \label{eq:meridian-vector-field}
\end{equation}
In a small angular window, the measurement-induced diffusion can be locally approximated by the reference process
\begin{equation}
    dq_t=\sqrt{2\epsilon}\,dW_t,
    \label{eq:local-brownian-reference}
\end{equation}
so that $D=2\epsilon$ and the transition kernel over time $T$ is
\begin{equation}
    K_T(q_0,q_T)
    =
    \frac{1}{\sqrt{4\pi\epsilon T}}
    \exp\left[-\frac{(q_T-q_0)^2}{4\epsilon T}\right].
    \label{eq:brownian-kernel}
\end{equation}
Prescribe two Gaussian endpoint distributions
\begin{equation}
    \mu_\nu=\mathcal N(m_\nu,s_\nu).
    \label{eq:gaussian-endpoints}
\end{equation}
Here $\nu=0,T$ labels the two endpoints, while $s_0$ and $s_T$ are variances.
The Sinkhorn endpoint coupling has the scaled form
\begin{equation}
    \pi^\ast(q_0,q_T)=a(q_0)K_T(q_0,q_T)b(q_T).
\end{equation}
The Gaussian nature of the problem enables us to directly finding the solution of the bridge problem as a joint Gaussian distribution with the correct marginal distributions.
For Gaussian marginals it is a Gaussian coupling with mean $(m_0,m_T)$ and covariance
\begin{equation}
    \Sigma_{0T}
    =
    \begin{pmatrix}
        s_0 & c\\
        c & s_T
    \end{pmatrix},
    \label{eq:gaussian-coupling}
\end{equation}
where the endpoint covariance is
\begin{equation}
    c=
    \frac{\sqrt{r^2+4s_0s_T}-r}{2}. \label{covar-c}
\end{equation}
The parameter appearing here is
\begin{equation}
    r=2\epsilon T .
\end{equation}
This is the explicit Sinkhorn solution for the endpoint problem.
Equivalently, the scaling functions $a$ and $b$ are exponentials of quadratic functions, and Eq.~(\ref{eq:gaussian-coupling}) is the corresponding Gaussian normal form - the explicit quadratic form of the exponents and the Gaussian integration directly leads to the needed covariance (\ref{covar-c}).

\begin{figure}
    \centering
    \includegraphics[width=\linewidth]{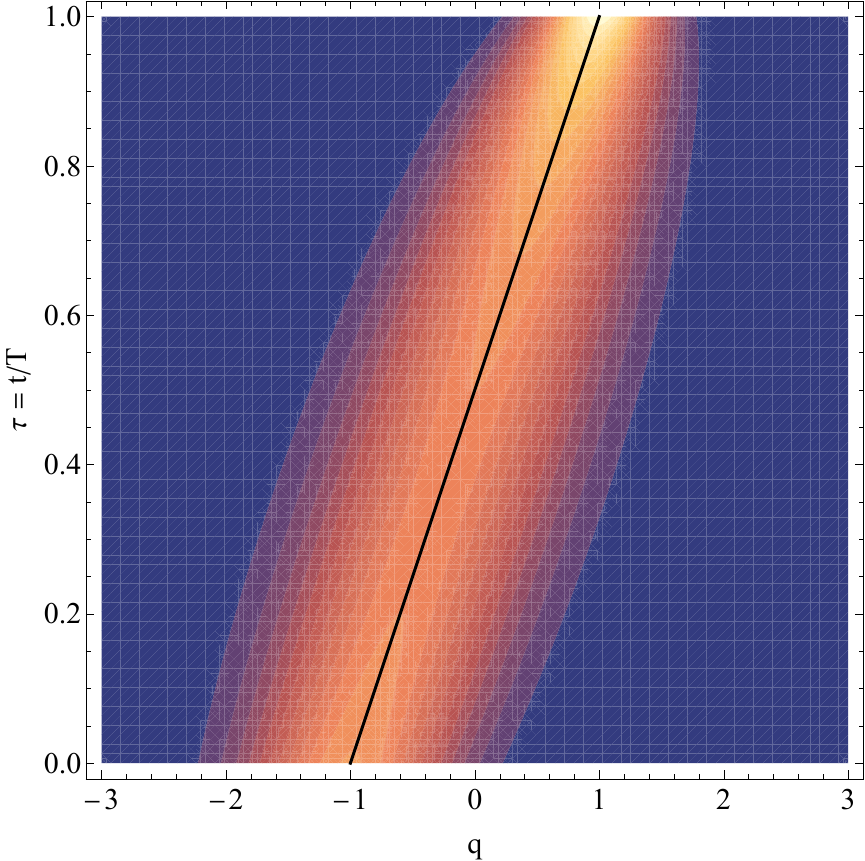}
    \caption{Illustration of the Gaussian bridge solution for the Sinkhorn bridge with a full terminal distribution.  
    The probability distribution is plotted as a density function of the coordinate $q$ and the scaled time $\tau = t/T$.  
    The marginal parameters are chosen as means of the initial and final marginals,
    $m_0 = -1, m_T = 1$, and covariance parameters $s_0 = 0.3, s_T = 0.1, c\approx 0.03,  r = 1$.
    The black line indicates the mean path $m_t$.
    }
    \label{fig:bridge}
\end{figure}

The whole bridge process is also Gaussian.
Writing $\tau=t/T$, its mean is
\begin{equation}
    m_t=(1-\tau)m_0+\tau m_T.
\end{equation}
The variance is
\begin{equation}
    s_t=(1-\tau)^2s_0+\tau^2s_T
    +2\tau(1-\tau)c
    +r\tau(1-\tau).
    \label{eq:gaussian-bridge-moments}
\end{equation}
The bridge solution is plotted in Fig.~\ref{fig:bridge}, showing a Gaussian probability distribution shifting its mean and variance in time in order to meet the initial and final boundary conditions.
The associated Fokker--Planck equation is
\begin{equation}
    \partial_t p^\ast
    =
    -\partial_q(b^\ast p^\ast)
    +
    \epsilon\,\partial_q^2p^\ast ,
\end{equation}
with optimal bridge drift
\begin{equation}
    b^\ast(q,t)
    =
    \dot m_t
    +
    \frac{\dot s_t-2\epsilon}{2s_t}
    (q-m_t).
    \label{eq:gaussian-bridge-drift}
\end{equation}
Since $D=2\epsilon$, the backward Sinkhorn potential satisfies
\begin{equation}
    b^\ast(q,t)=2\epsilon\,\partial_q\log\varphi(q,t).
    \label{eq:gaussian-potential-gradient}
\end{equation}
Thus the potential itself is quadratic in $q$:
\begin{equation}
    \partial_q\log\varphi(q,t)
    =
    \frac{\dot m_t}{2\epsilon}
    +
    \frac{\dot s_t-2\epsilon}{4\epsilon s_t}
    (q-m_t).
    \label{eq:gaussian-score-field}
\end{equation}

Because the present qubit realization has $f_A(q)=1$, the Sinkhorn control score along $A=\sigma_y/2$ is simply
\begin{equation}
    S_y(q,t)
    =
    \partial_q\log\varphi(q,t).
    \label{eq:gaussian-control-score}
\end{equation}
This is the full-distribution analogue of Eq.~(\ref{eq:qubit-score}).

We can now check the weak-value formula directly.
Let a backward effect or effect-like costate in the same meridian plane be
\begin{equation}
    E(t)=\alpha(t)I+\beta_x(t)\sigma_x+\beta_z(t)\sigma_z .
\end{equation}
Its likelihood on the chart is
\begin{equation}
    h_E(q,t)
    =
    \Tr[E(t)\rho(q)]
    =
    \alpha+\beta_x\sin q+\beta_z\cos q .
    \label{eq:meridian-effect-likelihood}
\end{equation}
A direct Pauli-matrix calculation gives
\begin{equation}
    2\Im
    \frac{\Tr[E(t)(\sigma_y/2)\rho(q)]}{\Tr[E(t)\rho(q)]}
    =
    \frac{\beta_x\cos q-\beta_z\sin q}
    {\alpha+\beta_x\sin q+\beta_z\cos q}.
    \label{eq:meridian-weak-value-check}
\end{equation}
This is equal to $\partial_q\log h_E(q,t)$, because $h_E(q,t)$ is given by Eq.~(\ref{eq:meridian-effect-likelihood}).
Thus, whenever the backward potential is represented by an effect likelihood $h_E$, the imaginary weak value is exactly the Sinkhorn score in Eq.~(\ref{eq:gaussian-control-score}).

For the Gaussian Sinkhorn potential in Eq.~(\ref{eq:gaussian-score-field}), a single qubit effect cannot represent the whole quadratic function globally, because $h_E(q,t)$ is trigonometric-affine in $q$.
However, the weak-value identification is exact pointwise.
At a point $(q_c,t)$ define
\begin{equation}
    \kappa_c(t)=\partial_q\log\varphi(q_c,t).
\end{equation}
Introduce the radial and tangent Pauli directions
\begin{equation}
    R_c=\sin q_c\,\sigma_x+\cos q_c\,\sigma_z.
\end{equation}
\begin{equation}
    T_c=\cos q_c\,\sigma_x-\sin q_c\,\sigma_z .
\end{equation}
The local effect
\begin{equation}
    E_c(t)=\lambda_c
    \left(I+a_c R_c+b_c T_c\right),
    \label{eq:local-effect-matching}
\end{equation}
where
\begin{equation}
    a_c=\frac{1-\kappa_c^2}{1+\kappa_c^2}.
\end{equation}
The other coefficient is
\begin{equation}
    b_c=\frac{2\kappa_c}{1+\kappa_c^2}.
\end{equation}
With an arbitrary positive scale $\lambda_c$, this effect satisfies
\begin{equation}
    \left.
    2\Im
    \frac{\Tr[E_c(t)(\sigma_y/2)\rho(q)]}{\Tr[E_c(t)\rho(q)]}
    \right|_{q=q_c}
    =
    \kappa_c(t)
    =
    S_y(q_c,t).
    \label{eq:local-weak-value-sinkhorn-match}
\end{equation}
Choosing $\lambda_c\le1/2$ makes $E_c$ a positive effect bounded by the identity; the overall scale does not affect the weak value.
In this sense, the full-distribution Sinkhorn potential supplies a pointwise family of backward effects, and their imaginary weak values reproduce the local bridge control field.

\subsection{Two Noncommuting Detectors on a Circle}

The preceding Gaussian example used a local Brownian approximation.
A physically explicit measurement model produces an exactly one-dimensional bridge on a circle.
Consider a qubit monitored continuously by two efficient detectors measuring the noncommuting observables $\sigma_x$ and $\sigma_z$, with measurement rates $\kappa_x$ and $\kappa_z$.  This situation was realized experimentally in the group of Irfan Siddiqi, where distributions of quantum trajectories were experimentally quantified \cite{hacohen2016quantum}.
Let the available Hamiltonian control be
\begin{equation}
    H(u)=\frac{u}{2}\sigma_y .
\end{equation}
Restricting to the $x$-$z$ plane of the Bloch sphere, write
\begin{equation}
    \rho=\frac{1}{2}(I+x\sigma_x+z\sigma_z).
\end{equation}
With the measurement-rate convention used here, the $x$ component of the It\^o equation is
\begin{equation}
    dx=
    \left(uz-\frac{\kappa_z}{2}x\right)dt
    +\sqrt{\kappa_x}(1-x^2)dW_x
    -\sqrt{\kappa_z}xz\,dW_z.
    \label{eq:two-detector-bloch-sde-x}
\end{equation}
The $z$ component is
\begin{equation}
    dz=
    \left(-ux-\frac{\kappa_x}{2}z\right)dt
    +\sqrt{\kappa_z}(1-z^2)dW_z
    -\sqrt{\kappa_x}xz\,dW_x .
    \label{eq:two-detector-bloch-sde-z}
\end{equation}
This is a simple physical realization of the state-space diffusion assumed above.
Parametrize the great circle by
\begin{equation}
    (x,z)=(\sin\theta,\cos\theta).
\end{equation}
This parametrization makes the $y$-Hamiltonian a translation of $\theta$.
Applying It\^o calculus gives
\begin{equation}
    \begin{split}
    d\theta
    ={}&
    \left[
    u+\frac{1}{2}(\kappa_x-\kappa_z)\sin\theta\cos\theta
    \right]dt
    \\
    &+
    \sqrt{\kappa_x}\cos\theta\,dW_x
    -\sqrt{\kappa_z}\sin\theta\,dW_z .
    \end{split}
    \label{eq:two-detector-angle-sde}
\end{equation}
Thus the probability density on $\theta\in[0,2\pi)$ obeys
\begin{equation}
    \partial_t p
    =
    -\partial_\theta[b(\theta,u)p]
    +\frac{1}{2}\partial_\theta^2[D(\theta)p],
    \label{eq:circle-fpe}
\end{equation}
where the drift is
\begin{equation}
    b(\theta,u)
    =
    u+\frac{1}{2}(\kappa_x-\kappa_z)\sin\theta\cos\theta.
\end{equation}
The scalar diffusion is
\begin{equation}
    D(\theta)
    =
    \kappa_x\cos^2\theta+\kappa_z\sin^2\theta .
    \label{eq:circle-drift-diffusion}
\end{equation}
See the discussion in Ref.~\cite{JordanSiddiqi2024} for more details.  Equivalently, the forward generator acting on test functions is
\begin{equation}
    \mathcal G f
    =
    b(\theta,u)\partial_\theta f
    +\frac{1}{2}D(\theta)\partial_\theta^2 f .
    \label{eq:circle-forward-generator}
\end{equation}
When $\kappa_x\gg \kappa_z$, the diffusion is largest near the $\sigma_z$ poles and the measurement tends to purify toward the $\sigma_x$ eigenstates; interchanging $x$ and $z$ gives the opposite limit.

For general $\kappa_x\neq\kappa_z$, Eqs.~(\ref{eq:circle-fpe})--(\ref{eq:circle-forward-generator}) give a periodic Sinkhorn problem with state-dependent diffusion.
The forward bridge potential obeys
\begin{equation}
    \partial_t\hat\varphi
    =
    \mathcal G^\dagger\hat\varphi.
\end{equation}
The backward bridge potential obeys
\begin{equation}
    \partial_t\varphi
    =
    -\mathcal G\varphi,
    \label{eq:circle-sinkhorn-system}
\end{equation}
with endpoint constraints
\begin{equation}
    p^\ast(\theta,t)
    =
    \frac{\hat\varphi(\theta,t)\varphi(\theta,t)}
    {\int_0^{2\pi}\hat\varphi(\theta,t)\varphi(\theta,t)\,d\theta}.
    \label{eq:circle-bridge-density}
\end{equation}
This is the continuous-time limit of Sinkhorn/IPFP on a periodic grid.

The equal-rate case $\kappa_x=\kappa_z=\kappa$ is analytic.
Then the angular stochastic equation becomes
\begin{equation}
    d\theta=u\,dt+\sqrt{\kappa}\,dW_t.
\end{equation}
The corresponding Fokker--Planck equation is
\begin{equation}
    \partial_t p=-u\partial_\theta p+\frac{\kappa}{2}\partial_\theta^2p .
\end{equation}
\begin{figure*}[t]
    \includegraphics[width=0.97\textwidth]{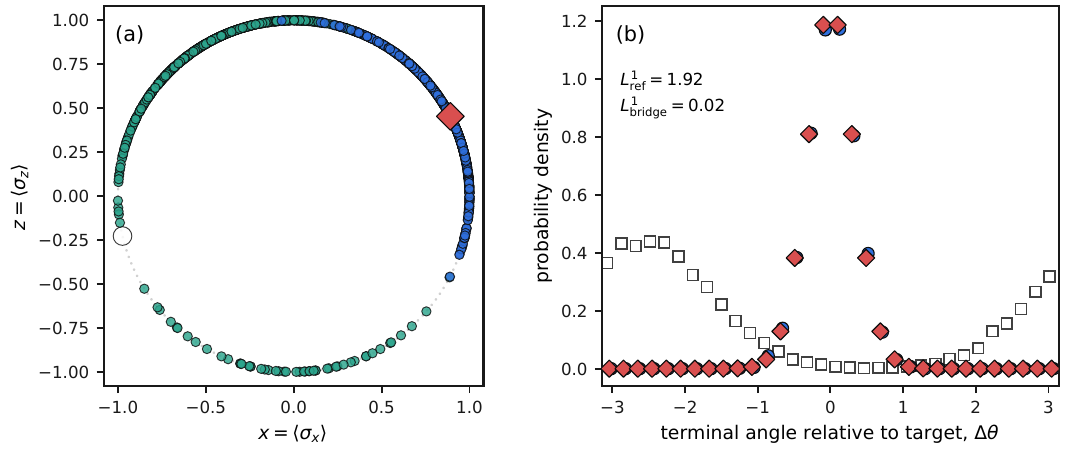}
    \caption{Numerical two-detector qubit bridge for equal measurement rates. Panel (a) shows sampled states on the $x$-$z$ Bloch circle: the white circle represents the initial state; green circles represent the controlled bridge at $t/T=1/2$; blue circles represent the controlled bridge at $t/T=1$; the red diamond represents the target mean. Points are samples and are not connected by trajectory lines. Panel (b) compares the terminal angle distributions relative to the target: gray squares represent the uncontrolled reference process; blue circles represent the process controlled by the weak-value/Sinkhorn score; red diamonds represent the prescribed target distribution. 
    The prescribed terminal distribution is a wrapped Gaussian in $\Delta\theta=\theta-\theta_T$ with mean $0$ and variance $0.10$, equivalently inverse variance, or concentration, $10$.
    Vertical bars show one-standard-deviation binomial sampling errors for the simulated distributions; most are smaller than the plotting symbols.
    The terminal $L^1$ distances from the target distribution are $L^1_{\rm ref}=1.92$ for the uncontrolled reference process and $L^1_{\rm bridge}=0.02$ for the weak-value/Sinkhorn-score-controlled process. Parameters are taken to be $T=1.5$, $\kappa_x=\kappa_z=0.55$, $u=0.25$, $\theta_i=-1.8$, and target concentration $10$.}
    \label{fig:two-detector-bridge}
\end{figure*}
Let $K_\tau(\theta,\theta')$ be the wrapped heat kernel with drift $u$:
\begin{equation}
    K_\tau(\theta,\theta')
    =
    \sum_{n\in\mathbb Z}
    \frac{
    \exp\!\left[
    -\frac{(\theta-\theta'-u\tau+2\pi n)^2}{2\kappa\tau}
    \right]}
    {\sqrt{2\pi\kappa\tau}} .
    \label{eq:wrapped-heat-kernel}
\end{equation}
For point boundary conditions $\theta(0)=\theta_i$ and $\theta(T)=\theta_f$,
\begin{equation}
    \hat\varphi(\theta,t)=K_t(\theta,\theta_i).
    \label{eq:circle-delta-potentials}
\end{equation}
The backward potential is
\begin{equation}
    \varphi(\theta,t)=K_{T-t}(\theta_f,\theta).
\end{equation}
More general endpoint distributions are obtained by integrating this kernel against the corresponding Sinkhorn scaling functions.
The Doob-corrected bridge drift is
\begin{equation}
    b^\ast(\theta,t)
    =
    u+\kappa\,\partial_\theta\log\varphi(\theta,t).
    \label{eq:circle-doob-drift}
\end{equation}
When a single winding sector dominates the wrapped kernel,
\begin{equation}
    \partial_\theta\log\varphi(\theta,t)
    \simeq
    \frac{\theta_f-\theta-u(T-t)+2\pi n_\ast}
    {\kappa(T-t)} ,
    \label{eq:circle-dominant-winding-score}
\end{equation}
so Eq.~(\ref{eq:circle-doob-drift}) reduces to the familiar Brownian-bridge steering term on the appropriate lift of the circle.

Finally, the weak-value interpretation is immediate.
Because $A=\sigma_y/2$ generates translations of $\theta$,
\begin{equation}
    f_A(\theta)=1.
\end{equation}
The control score along this direction is therefore
\begin{equation}
    S_y(\theta,t)=\partial_\theta\log\varphi(\theta,t).
    \label{eq:circle-control-score}
\end{equation}
Whenever the local backward potential can be represented as an effect likelihood $h_E(\theta,t)=\Tr[E(t)\rho(\theta)]$, this score is
\begin{equation}
    S_y(\theta,t)
    =
    2\Im
    \frac{\Tr[E(t)(\sigma_y/2)\rho(\theta)]}
    {\Tr[E(t)\rho(\theta)]}.
    \label{eq:circle-weak-value-score}
\end{equation}
Figure~\ref{fig:two-detector-bridge} demonstrates the corresponding bridge-control synthesis.
We compute the terminal Sinkhorn scaling for a prescribed target distribution on the circle and simulate the controlled process with drift $u+\kappa S_y(\theta,t)$.
The bridge ensemble is steered to the desired terminal distribution, while the uncontrolled reference ensemble is not.
Thus the two-detector model gives a concrete noncommuting-measurement realization of the same statement: the Sinkhorn potential sets the Doob drift of the conditioned diffusion, and its Hamiltonian directional derivative is the imaginary weak value.
The $L_1$ distance from the target distribution is reduced from $1.92$ for the uncontrolled reference process to $0.02$ for the Sinkhorn-score-controlled process, illustrating the practical importance of these results to quantum control.

\section{Outlook}
\label{sec:outlook}
Our framework suggests a practical pipeline.
First, choose a reference continuously monitored dynamics and discretize its induced state-space Fokker--Planck kernel.
Second, impose either endpoint distributions or a terminal effect and compute the Sinkhorn/Doob potential.
Third, convert the potential into local weak-value scores along available Hamiltonian and measurement-control directions.
Finally, synthesize a feedback protocol by projecting the ideal bridge drift into the available control algebra.
Operationally, once the final boundary condition has been encoded as a backward effect or a Sinkhorn potential, the imaginary weak values of the available control generators are the local logarithmic responses to those controls.
They therefore provide a direct feedback prescription: compute these imaginary weak values and use the resulting scores as the steering field toward the desired terminal boundary.
For a quadratic-cost feedback law, the optimal infinitesimal control is obtained by projecting these scores onto the available control directions.
For bounded controls, their signs become switching functions that determine which control extremum is selected.

The conceptual payoff is that weak measurement supplies a natural entropic regularization of quantum state transport.
The Sinkhorn potential is not only a computational scaling factor.
In quantum state space, it is the backward likelihood of a future measurement boundary, and its physically accessible directional derivatives are imaginary weak values.

\begin{acknowledgments}
MO was supported by the Cross-ministerial Strategic Innovation Promotion Program (SIP) of the Cabinet Office, Government of Japan (No. 23836436).
ANJ thanks Olga Movilla Miangolarra, Ralph Sabbagh, and Tryphon T. Georgiou for helpful discussions on Schr\"odinger bridges.  ANJ's research was supported by a grant from the John Templeton Foundation (Grant \# 63209).
\end{acknowledgments}

\bibliography{references}

@article{aharonov1964time,
  title={Time symmetry in the quantum process of measurement},
  author={Aharonov, Yakir and Bergmann, Peter G and Lebowitz, Joel L},
  journal={Physical Review},
  volume={134},
  number={6B},
  pages={B1410},
  year={1964},
  publisher={APS}
}

@article{hubner1992explicit,
  title={Explicit computation of the Bures distance for density matrices},
  author={H{\"u}bner, Matthias},
  journal={Physics Letters A},
  volume={163},
  number={4},
  pages={239--242},
  year={1992},
  publisher={Elsevier}
}

@article{bures1969extension,
  title={An extension of Kakutani’s theorem on infinite product measures to the tensor product of semifinite $w^\ast$-algebras},
  author={Bures, Donald},
  journal={Transactions of the American Mathematical Society},
  volume={135},
  pages={199--212},
  year={1969}
}

@article{guevara2015quantum,
  title={Quantum state smoothing},
  author={Guevara, Ivonne and Wiseman, Howard},
  journal={Physical review letters},
  volume={115},
  number={18},
  pages={180407},
  year={2015},
  publisher={APS}
}

@article{zhang2017prediction,
  title={Prediction and retrodiction with continuously monitored Gaussian states},
  author={Zhang, Jinglei and M{\o}lmer, Klaus},
  journal={Physical Review A},
  volume={96},
  number={6},
  pages={062131},
  year={2017},
  publisher={APS}
}

@article{garcia2017past,
  title={Past observable dynamics of a continuously monitored qubit},
  author={Garc{\'\i}a-Pintos, Luis Pedro and Dressel, Justin},
  journal={Physical Review A},
  volume={96},
  number={6},
  pages={062110},
  year={2017},
  publisher={APS}
}

@article{weber2014mapping,
  title={Mapping the optimal route between two quantum states},
  author={Weber, SJ and Chantasri, Areeya and Dressel, Justin and Jordan, Andrew N and Murch, Kater W and Siddiqi, Irfan},
  journal={Nature},
  volume={511},
  number={7511},
  pages={570--573},
  year={2014},
  publisher={Nature Publishing Group UK London}
}

@article{hacohen2016quantum,
  title={Quantum dynamics of simultaneously measured non-commuting observables},
  author={Hacohen-Gourgy, Shay and Martin, Leigh S and Flurin, Emmanuel and Ramasesh, Vinay V and Whaley, K Birgitta and Siddiqi, Irfan},
  journal={Nature},
  volume={538},
  number={7626},
  pages={491--494},
  year={2016},
  publisher={Nature Publishing Group UK London}
}

@article{boscain2021introduction,
  title={Introduction to the Pontryagin maximum principle for quantum optimal control},
  author={Boscain, Ugo and Sigalotti, Mario and Sugny, Dominique},
  journal={PRX Quantum},
  volume={2},
  number={3},
  pages={030203},
  year={2021},
  publisher={APS}
}

@article{chantasri2015stochastic,
  title={Stochastic path-integral formalism for continuous quantum measurement},
  author={Chantasri, Areeya and Jordan, Andrew N},
  journal={Physical Review A},
  volume={92},
  number={3},
  pages={032125},
  year={2015},
  publisher={APS}
}

@article{idel2016review,
  title={A review of matrix scaling and Sinkhorn's normal form for matrices and positive maps},
  author={Idel, Martin},
  journal={arXiv preprint arXiv:1609.06349},
  year={2016}
}

@book{schleich2015quantum,
  title={Quantum optics in phase space},
  author={Schleich, Wolfgang P},
  year={2015},
  publisher={John Wiley \& Sons}
}

@article{arvidsson2024properties,
  title={Properties and applications of the Kirkwood--Dirac distribution},
  author={Arvidsson-Shukur, David RM and Braasch Jr, William F and De Bievre, Stephan and Dressel, Justin and Jordan, Andrew N and Langrenez, Christopher and Lostaglio, Matteo and Lundeen, Jeff S and Halpern, Nicole Yunger},
  journal={New Journal of Physics},
  volume={26},
  number={12},
  pages={121201},
  year={2024},
  publisher={IOP Publishing}
}

@article{hacohen2018incoherent,
  title={Incoherent qubit control using the quantum Zeno effect},
  author={Hacohen-Gourgy, Shay and Garc{\'\i}a-Pintos, Luis Pedro and Martin, Leigh S and Dressel, Justin and Siddiqi, Irfan},
  journal={Physical review letters},
  volume={120},
  number={2},
  pages={020505},
  year={2018},
  publisher={APS}
}

@article{lewalle2024optimal,
  title={Optimal zeno dragging for quantum control: a shortcut to zeno with action-based scheduling optimization},
  author={Lewalle, Philippe and Zhang, Yipei and Whaley, K Birgitta},
  journal={PRX Quantum},
  volume={5},
  number={2},
  pages={020366},
  year={2024},
  publisher={APS}
}

@article{kokaew2026quantum,
  title={Quantum state preparation control in noisy environment via most-likely paths},
  author={Kokaew, Wirawat and Chotibut, Thiparat and Chantasri, Areeya},
  journal={Quantum Information Processing},
  volume={25},
  number={1},
  pages={26},
  year={2026},
  publisher={Springer}
}

@article{carlini2006time,
  title={Time-optimal quantum evolution},
  author={Carlini, Alberto and Hosoya, Akio and Koike, Tatsuhiko and Okudaira, Yosuke},
  journal={Physical review letters},
  volume={96},
  number={6},
  pages={060503},
  year={2006},
  publisher={APS}
}

@article{guery2019shortcuts,
  title={Shortcuts to adiabaticity: Concepts, methods, and applications},
  author={Gu{\'e}ry-Odelin, David and Ruschhaupt, Andreas and Kiely, Anthony and Torrontegui, Erik and Mart{\'\i}nez-Garaot, Sofia and Muga, Juan Gonzalo},
  journal={Reviews of Modern Physics},
  volume={91},
  number={4},
  pages={045001},
  year={2019},
  publisher={APS}
}

@article{Ohzeki2026TwoBoundary,
  title={Coherent Quantum Schrodinger Bridge: Two-Boundary Optimal Control for Quantum Algorithm Design},
  author={Ohzeki, Masayuki},
  journal={arXiv preprint arXiv:2607.10550},
  year={2026}
}

@book{doob1984classical,
  title={Classical potential theory and its probabilistic counterpart},
  author={Doob, Joseph L and others},
  volume={19},
  year={1984},
  publisher={Springer}
}

@article{sinkhorn1964relationship,
  title={A relationship between arbitrary positive matrices and doubly stochastic matrices},
  author={Sinkhorn, Richard},
  journal={The annals of mathematical statistics},
  volume={35},
  number={2},
  pages={876--879},
  year={1964},
  publisher={JSTOR}
}

@article{georgiou2015positive,
  title={Positive contraction mappings for classical and quantum Schr{\"o}dinger systems},
  author={Georgiou, Tryphon T and Pavon, Michele},
  journal={Journal of Mathematical Physics},
  volume={56},
  number={3},
  year={2015},
  publisher={AIP Publishing}
}

@article{Aharonov1988,
  title = {How the Result of a Measurement of a Component of the Spin of a Spin-1/2 Particle Can Turn Out to Be 100},
  author = {Aharonov, Yakir and Albert, David Z. and Vaidman, Lev},
  journal = {Phys. Rev. Lett.},
  volume = {60},
  pages = {1351--1354},
  year = {1988},
  doi = {10.1103/PhysRevLett.60.1351}
}

@article{Schrodinger1931,
  title = {{\"U}ber die Umkehrung der Naturgesetze},
  author = {Schr{\"o}dinger, Erwin},
  journal = {Sitzungsberichte der Preussischen Akademie der Wissenschaften, Physikalisch-mathematische Klasse},
  volume = {144},
  pages = {144--153},
  year = {1931}
}

@article{HuJordan2023Driving,
  title = {Quantum State Driving along Arbitrary Trajectories},
  author = {Hu, Le and Jordan, Andrew N.},
  journal = {Phys. Rev. Research},
  volume = {5},
  pages = {033045},
  year = {2023},
  doi = {10.1103/PhysRevResearch.5.033045},
  eprint = {2211.02457},
  archivePrefix = {arXiv},
  primaryClass = {quant-ph}
}

@article{HuJordan2026Collapse,
  title = {Describing the Wave Function Collapse Process with a State-dependent Hamiltonian},
  author = {Hu, Le and Jordan, Andrew N.},
  journal = {Quantum Studies: Mathematics and Foundations},
  volume = {13},
  pages = {22},
  year = {2026},
  eprint = {2301.09274},
  archivePrefix = {arXiv},
  primaryClass = {quant-ph}
}

@article{KarmakarJordan2026CDJPontryagin,
  title = {{CDJ}-Pontryagin Optimal Control for General Continuously Monitored Quantum Systems},
  author = {Karmakar, Tathagata and Jordan, Andrew N.},
  journal = {Quantum},
  volume = {10},
  pages = {2043},
  year = {2026},
  doi = {10.22331/q-2026-03-24-2043},
  eprint = {2504.08173},
  archivePrefix = {arXiv},
  primaryClass = {quant-ph}
}

@article{ChantasriDresselJordan2013,
  title = {Action Principle for Continuous Quantum Measurement},
  author = {Chantasri, Areeya and Dressel, Justin and Jordan, Andrew N.},
  journal = {Phys. Rev. A},
  volume = {88},
  pages = {042110},
  year = {2013},
  doi = {10.1103/PhysRevA.88.042110}
}

@article{DresselJordan2012,
  title = {Significance of the Imaginary Part of the Weak Value},
  author = {Dressel, Justin and Jordan, Andrew N.},
  journal = {Phys. Rev. A},
  volume = {85},
  pages = {012107},
  year = {2012},
  doi = {10.1103/PhysRevA.85.012107},
  eprint = {1112.3986},
  archivePrefix = {arXiv},
  primaryClass = {quant-ph}
}

@article{Dressel2014,
  title = {Colloquium: Understanding Quantum Weak Values: Basics and Applications},
  author = {Dressel, Justin and Malik, Mehul and Miatto, Filippo M. and Jordan, Andrew N. and Boyd, Robert W.},
  journal = {Rev. Mod. Phys.},
  volume = {86},
  pages = {307--316},
  year = {2014},
  doi = {10.1103/RevModPhys.86.307}
}

@article{MovillaSabbaghGeorgiou2025,
  title = {Quantum {Schr\"odinger} Bridges: Large Deviations and Time-Symmetric Ensembles},
  author = {Movilla Miangolarra, Olga and Sabbagh, Ralph and Georgiou, Tryphon T.},
  journal = {Phys. Rev. A},
  volume = {112},
  pages = {012202},
  year = {2025},
  doi = {10.1103/k35b-rkct},
  eprint = {2503.05886},
  archivePrefix = {arXiv},
  primaryClass = {quant-ph}
}

@article{Leonard2014Survey,
  title = {A Survey of the Schr\"odinger Problem and Some of Its Connections with Optimal Transport},
  author = {L{\'e}onard, Christian},
  journal = {Discrete and Continuous Dynamical Systems},
  volume = {34},
  number = {4},
  pages = {1533--1574},
  year = {2014},
  doi = {10.3934/dcds.2014.34.1533}
}

@article{ChenGeorgiouPavon2016,
  title = {Optimal Steering of a Linear Stochastic System to a Final Probability Distribution, Part {I}},
  author = {Chen, Yongxin and Georgiou, Tryphon T. and Pavon, Michele},
  journal = {IEEE Transactions on Automatic Control},
  volume = {61},
  number = {5},
  pages = {1158--1169},
  year = {2016},
  doi = {10.1109/TAC.2015.2457784}
}

@article{Sinkhorn1967,
  title = {Diagonal Equivalence to Matrices with Prescribed Row and Column Sums},
  author = {Sinkhorn, Richard},
  journal = {The American Mathematical Monthly},
  volume = {74},
  number = {4},
  pages = {402--405},
  year = {1967},
  doi = {10.2307/2314570}
}

@article{PeyreCuturi2019,
  title = {Computational Optimal Transport},
  author = {Peyr{\'e}, Gabriel and Cuturi, Marco},
  journal = {Foundations and Trends in Machine Learning},
  volume = {11},
  number = {5--6},
  pages = {355--607},
  year = {2019},
  doi = {10.1561/2200000073}
}

@book{JordanSiddiqi2024,
  title = {Quantum Measurement: Theory and Practice},
  author = {Jordan, Andrew N. and Siddiqi, Irfan A.},
  publisher = {Cambridge University Press},
  address = {Cambridge},
  year = {2024},
  doi = {10.1017/9781009103909},
  isbn = {9781009103909}
}

@book{Jacobs2014,
  title = {Quantum Measurement Theory and Its Applications},
  author = {Jacobs, Kurt},
  publisher = {Cambridge University Press},
  address = {Cambridge},
  year = {2014},
  doi = {10.1017/CBO9781139179027}
}

@article{Minev2019Nature,
   author = {Z. K. Minev and S. O. Mundhada and S. Shankar and P. Reinhold and R. Gutiérrez-Jáuregui and R. J. Schoelkopf and M. Mirrahimi and H. J. Carmichael and M. H. Devoret},
   doi = {10.1038/s41586-019-1287-z},
   issn = {0028-0836},
   issue = {7760},
   journal = {Nature},
   month = {6},
   pages = {200-204},
   title = {To catch and reverse a quantum jump mid-flight},
   volume = {570},
   url = {http://arxiv.org/abs/1803.00545 http://www.nature.com/articles/s41586-019-1287-z},
   year = {2019},
}

@article{Murch2013,
   author = {K. W. Murch and S. J. Weber and C. Macklin and I. Siddiqi},
   doi = {10.1038/nature12539},
   issn = {0028-0836},
   issue = {7470},
   journal = {Nature},
   month = {10},
   pages = {211-214},
   title = {Observing single quantum trajectories of a superconducting quantum bit},
   volume = {502},
   year = {2013},
}

@article{Weber2014,
   author = {S. J. Weber and A. Chantasri and J. Dressel and A. N. Jordan and K. W. Murch and I. Siddiqi},
   doi = {10.1038/nature13559},
   issn = {0028-0836},
   issue = {7511},
   journal = {Nature},
   month = {7},
   pages = {570-573},
   title = {Mapping the optimal route between two quantum states},
   volume = {511},
   year = {2014},
}

@article{Rabi_vijay,
  title = {Stabilizing Rabi oscillations in a superconducting qubit using quantum feedback},
  volume = {490},
  ISSN = {1476-4687},
  url = {http://dx.doi.org/10.1038/nature11505},
  DOI = {10.1038/nature11505},
  number = {7418},
  journal = {Nature},
  publisher = {Springer Science and Business Media LLC},
  author = {Vijay,  R. and Macklin,  C. and Slichter,  D. H. and Weber,  S. J. and Murch,  K. W. and Naik,  R. and Korotkov,  A. N. and Siddiqi,  I.},
  year = {2012},
  month = oct,
  pages = {77–80}
}

@article{patti2017linear,
  title={Linear feedback stabilization of a dispersively monitored qubit},
  author={Patti, Taylor Lee and Chantasri, Areeya and Garc{\'\i}a-Pintos, Luis Pedro and Jordan, Andrew N and Dressel, Justin},
  journal={Physical Review A},
  volume={96},
  number={2},
  pages={022311},
  year={2017},
  publisher={APS}
}

@article{QEC_livingston,
  title = {Experimental demonstration of continuous quantum error correction},
  volume = {13},
  ISSN = {2041-1723},
  url = {http://dx.doi.org/10.1038/s41467-022-29906-0},
  DOI = {10.1038/s41467-022-29906-0},
  number = {1},
  journal = {Nature Communications},
  publisher = {Springer Science and Business Media LLC},
  author = {Livingston,  William P. and Blok,  Machiel S. and Flurin,  Emmanuel and Dressel,  Justin and Jordan,  Andrew N. and Siddiqi,  Irfan},
  year = {2022},
  month = apr 
}

\end{document}